\documentclass[manuscript,nonacm]{acmart} 
\AtBeginDocument{%
  }

\setcopyright{acmlicensed}
\copyrightyear{2018}
\acmYear{2018}
\acmDOI{XXXXXXX.XXXXXXX}
\acmConference[Conference acronym 'XX]{Make sure to enter the correct
  conference title from your rights confirmation email}{June 03--05,
  2018}{Woodstock, NY}
\acmISBN{978-1-4503-XXXX-X/2018/06}

\usepackage{caption}
\usepackage{booktabs}   
\usepackage{multirow}   
\usepackage{graphicx}   
\usepackage{amsmath} 
\usepackage{subcaption}
\usepackage{url}

\begin{document}

\title{Making Absence Visible: The Roles of Reference and Prompting in Recognizing Missing Information
}

\author{Hagit Ben Shoshan}
\email{hagitphd@gmail.com}
\orcid{0009-0007-5945-695X} 
\affiliation{%
  \institution{University of Haifa}
  \city{Haifa}
  \country{Israel}}
 
\author{Joel Lanir}
\orcid{0000-0002-9838-5142}
\affiliation{%
  \institution{University of Haifa}
  \city{Haifa}
  \country{Israel}}
\email{jlanir@is.haifa.ac.il}

\author{Pavel Goldstein}
\orcid{0000-0002-5224-1725}
\affiliation{%
  \institution{University of Haifa}
  \city{Haifa}
  \country{Israel}}
\email{pavelg@stat.haifa.ac.il} 

\author{Osnat Mokryn}
\orcid{0000-0002-1241-9015}
\affiliation{%
  \institution{University of Haifa}
  \city{Haifa}
  \country{Israel}}
\email{omokryn@is.haifa.ac.il}

\renewcommand{\shortauthors}{Ben Shoshan et al.}

\begin{abstract}
 Interactive systems that explain data, or support decision making often emphasize what is present while overlooking what is expected but missing. This presence bias limits users’ ability to form complete mental models of a dataset or situation. Detecting absence depends on expectations about what should be there, yet interfaces rarely help users form such expectations. We present an experimental study examining how reference framing and prompting influence people’s ability to recognize expected but missing categories in datasets. Participants compared distributions across three domains (energy, wealth, and regime) under two reference conditions: Global, presenting a unified population baseline, and Partial, showing several concrete exemplars. Results indicate that absence detection was higher with Partial reference than with Global reference, suggesting that partial, samples-based framing can support expectation formation and absence detection. When participants were prompted to look for what was missing, absence detection rose sharply. We discuss implications for interactive user interfaces and expectation-based visualization design, while considering cognitive trade-offs of reference structures and guided attention.
\end{abstract}



\received{20 February 2007}
\received[revised]{12 March 2009}
\received[accepted]{5 June 2009}

\maketitle

\section{Introduction}

\begin{quote}
\textit{Gregory:} ``Is there any other point to which you would wish to draw my attention?'' \\
\textit{Holmes:} ``To the curious incident of the dog in the night-time.'' \\
\textit{Gregory:} ``The dog did nothing in the night-time.'' \\
\textit{Holmes:} ``That was the curious incident.'' \\
\hfill---\textit{Arthur Conan Doyle, ``Silver Blaze'' (1892)}~\cite{doyle2024adventure}
\end{quote}

Absences are often as revealing as presences. Arthur Conan Doyle’s \textit{Silver Blaze} dramatizes this through Sherlock Holmes’s attention to the dog that did not bark. Under ordinary circumstances, a watchdog is \emph{expected} to bark; its silence was therefore not emptiness but a violation of expectation—a missing event that became a decisive clue. The case illustrates a general principle: absence is meaningful only against a model of what should have been present. It is not a signal on its own, but a deviation from expectation~\cite{farennikova2013seeing,martin2013seeing,cavedon2017touching}.  

Expectation frames govern both everyday inference and data interpretation. We notice a friend who fails to arrive, or a skipped heartbeat, precisely because we anticipated them. In data analysis and visualization, insight arises not only from observed values but also from those that are \emph{expected yet missing}. A product review that omits any mention of ``price'' or a national dataset lacking ``healthcare'' may reveal more by what it leaves out than by what it shows, provided the observer has a reference of what is typical.

Human cognition, however, rarely performs this comparison automatically. The well-documented \textit{feature-positive effect}~\cite{newman1980feature} shows that people detect present cues more readily than absent ones and often overlook missing information unless explicitly guided~\cite{hearst1989backward,mazor2020distinct}. Judgments of absence are slower, less confident, and strongly dependent on contextual support~\cite{tversky1974judgment,legrenzi1993focussing}. In cognitive terms, absence is perceived when an internal prediction is violated. In interface terms, it becomes visible only when expectations are made explicit.  

This expectation-based perspective reframes a classic visualization problem. Standard charts excel at showing what exists, but provide few cues for what is missing and should exist. A missing bar may be indistinguishable from an unmeasured one; a gap may simply disappear into the background. Prior work in visualization and decision-making shows that comparative structures, such as baselines, benchmarks, and base rates, make expectations explicit and enable more accurate reasoning~\cite{attfield2010sense,margoni2024violation,gigerenzer1988presentation}. From this standpoint, reference frames function as \textit{externalized expectations} that allow viewers to detect deviations, including absences. Computational methods such as Latent Personal Analysis (LPA) and Learning via Surprisability (LvS) formalize this idea by building population-level reference models that expose deviations from expectation~\cite{mokryn2021domain,mokryn2025interpretable}. LvS then identifies \textit{missing commonalities} as features prevalent in the population but absent in a specific instance, thus turning \textit{absence} into structured information.  

Here, we examine how reference framing and guidance jointly influence the recognition of missing information in data visualizations. 
We hypothesized that partial reference frames would encourage detection of absence by prompting viewers to internally complete missing context. We further hypothesized that explicit prompting, which provides guided attentional support, would substantially increase detection across both reference conditions.

Using open-ended responses to comparative bar-chart visualizations, we measure how people notice missing categories under (1) partial exemplar reference and (2) global aggregate reference, both before and after guided prompting.

Our contributions are threefold:
\begin{enumerate}
    \item We introduce the concept of \textit{expectation-based visualization}, which embeds explicit reference models to externalize what ``should be'' present.
    \item We provide empirical evidence that reference framing meaningfully shapes how people recognize missing information: partial frames prompt viewers to reconstruct expectations, whereas global frames support direct comparison. Guided prompting dramatically amplifies detection across both conditions, showing that attentional guidance can overcome the inherent presence bias in spontaneous interpretation.
    \item We outline design implications for intelligent visualization systems that make expectations explicit and guide users’ attention to missing information, balancing cognitive effort with insight on how absences are presented.
\end{enumerate}

By connecting cognitive bias (the feature-positive effect) with interactive design, we show how visualization systems can help humans ``see what is not there''. 
Absences, when framed through explicit expectations and guided attention, become perceptible, interpretable, and actionable - but only when users are prompted to adopt an expectation-oriented stance.

\section{Related Work}
We review key cognitive principles and empirical findings that demonstrate how human perception and reasoning are systematically biased towards detecting presence rather than absence, highlighting the crucial role of contextual expectations and internal models in enabling the detection of missing features.

Philosophy offers a complementary view. Mumford \cite{mumford2021absence} argues that absence is neither directly perceived nor merely inferred. Instead, it is a \textit{hybrid phenomenon}: our senses register what is present, our minds compare this against what is expected, and the result is a felt experience of absence, a kind of ``user illusion''. This account suggests that absence perception is itself a reference-framing process: without an expectation against which the present is compared, no absence would be experienced. 

\subsection{Presence bias}
Presence bias is the systematic tendency to detect, learn, and reason from things that are present more readily than from things that are absent. In learning and judgment, this asymmetry is well known as the feature-positive effect (FPE): organisms more easily associate ``A and B'' than ``A and not B'', and people process present cues more fluently than missing ones~\cite{newman1980feature}. Evidence for this bias appears early in development. Infants attend more to added than to removed features, revealing a deep perceptual asymmetry~\cite{coldren2000asymmetries}. As Kahneman~\cite{kahneman2011thinking} notes, people construct coherent stories from available cues and tend to ignore what is missing unless it is made salient.

In data visualization, this cognitive bias has concrete implications. Human vision is tuned to presence over absence: targets defined by an added feature ``pop out'', whereas those defined by a missing feature are systematically harder to detect—the classic search asymmetry~\cite{treisman1988feature}. In quantitative displays, zeros are particularly meaningful because they represent true absences rather than missing or incomplete data. Making such zeros explicit is therefore crucial to counteract presence bias: otherwise, attention focuses only on where something happens, not where it does not. When data values are absent, viewers often fail to perceive that feature at all, interpreting the omission as missing information rather than a legitimate zero. Research on visualizing missing data shows that explicit markers or gaps improve comprehension and decision-making~\cite{song2018s}. 

\subsection{Absence and missing data}
Absences are not merely missing data, but a missing part of a cognitive reference frame, features that are to be expected, yet are missing. Absence is not a data-quality issue but an expectation-dependent cognitive phenomenon. Unlike absence, missing values refer to data quality issues. In prior work on visualizing missing data, Kandel \cite{kandel2011wrangler} addresses how systems can surface missing values such as nulls, incomplete rows, or gaps arising from data collection. In contrast, our study examines cases where the data themselves are complete, but the viewer fails to notice a category that is semantically expected, given a reference frame.  Our focus is on how reference structures and attentional guidance influence the recognition of  \textbf{absent expected categories} that ``should have been present'', a capability fundamentally different from detecting missing entries in the data.
 The experimental study reported here applies this reasoning to data visualization, testing how different forms of reference framing (global versus partial) shape human detection of missing commonalities in graphical data.

\subsection{Perceiving and Detecting Absence}

\paragraph{Cognitive and neural perspectives of absence.} Experiences of absence are common, such as noticing that a car is missing from its usual place. Yet, the neural and cognitive mechanisms underlying how the mind ``perceives nothing'' remain debated~\cite{mumford2021absence,barnett2024symbolic}. Some theories argue that absence can be directly perceived, while others maintain it is inferred from contextual cues~\cite{cavedon2017touching,mumford2021absence}. The distinction between perceiving and inferring absence is subtle but fundamental: it determines whether absence is experienced as a percept or as a cognitive reconstruction.

\paragraph{The role of internal models.}
Recognizing absence requires prior expectations. Without anticipating a particular feature, one cannot register its lack. Human perception continually compares incoming sensory input against internal predictive models; absence is detected when expected signals fail to appear. Thus, the perception of absence arises from a mismatch between prediction and observation. As Mazor~\cite{mazor2025inference} notes, such inference depends on a self-model capable of transforming the ``absence of evidence'' into ``evidence of absence''.

Research using the violation-of-expectation (VOE) paradigm supports this model: when expected objects or events are missing, infants exhibit surprise, prompting learning and model revision~\cite{margoni2024violation}. This demonstrates that absence detection is not a passive process but an active one: triggering attention, expectation adjustment, and cognitive updating.

\paragraph{Absence detection as cognitive inference }
Detecting absence involves comparing current input with internalized expectations and reconciling discrepancies between them. When the observed and the expected diverge, attention shifts and higher-order inference processes engage to explain the gap. Recognition of absence thus depends on the same predictive mechanisms that support surprise and learning.
Empirical work confirms that judgments of absence differ systematically from those of presence: people are slower and less confident when deciding that a stimulus is absent, even when correct~\cite{meuwese2014subjective,mazor2020distinct}. Absence detection is therefore a dynamic and cognitively demanding process—less fluent, slower, and less confident than presence detection, yet essential for adaptive learning and reasoning about what is missing.
\subsection{Reference Framing and the Role of Expectations}
A recurring insight across cognitive psychology and visual analytics is that the ability to detect absence depends on the presence of an explicit or internalized reference frame. Gigerenzer et al.~\cite{gigerenzer1988presentation} demonstrated that base-rate use is not a fixed heuristic bias but a function of \textit{problem representation}: when participants directly observed the random-sampling process rather than merely hearing about it, their probability judgments approached Bayesian reasoning. Presentation, therefore, shapes the internal model against which new evidence is evaluated. The same logic underlies the \textit{violation-of-expectation} paradigm in developmental research~\cite{margoni2024violation}, where infants’ surprise at impossible or missing events reveals that perception operates through predictive mental models—an absence is noticed only when it violates an expectation of presence.
In visual reasoning, Attfield et al.~\cite{attfield2010sense} extend this principle to analytic contexts, describing sensemaking as an iterative loop between foraging for data and updating explanatory hypotheses. Effective visualization systems, they argue, are those that externalize expectations, enabling users to recognize both anomalies and absences through visible discrepancies between what is observed and what is assumed. 
Neisser’s perceptual cycle account offers a complementary cognitive mechanism for how reference frames support absence detection: perception is guided by anticipatory schemata that direct attention and are updated through sampling. When the available input is partial, viewers must actively construct and refine an internal model of ``what should be there” \cite{neisser1976cognition}

Together, these traditions converge on the idea that \textit{reference framing} provides the cognitive substrate for absence detection: it anchors perception to an explicit baseline, making deviations—positive or negative—computationally and perceptually tractable.
From a computational perspective, this logic is mirrored in methods that formalize reference-based comparison as a means of expectation modeling. Approaches such as predictive coding, surprisal-based modeling, and distributional reference frameworks treat data interpretation as a process of quantifying deviation from an expected baseline~\cite{mokryn2021domain,feng2024holens}. In visual analytics, population-level summaries (averages, density plots, or global baselines) often function as reference frames, contextualizing local data points and exposing deviations or absences. This design principle is consistent with approaches in recent work that compare individual observations to model or aggregate expectations~\cite{suschnigg2025mandala,mokryn2025interpretable}. Whether implemented through probabilistic inference, divergence metrics, or learned priors, these methods instantiate the same principle that underlies human reasoning about absence: the identification of what is missing emerges only through explicit contrast with what was expected to be there. In the context of data visualization, reference framing thus functions as a bridge between cognitive and computational reasoning about expectation, clarifying how context shapes the salience of absence.

\section{Hypotheses}
The detection of missing information depends on expectations: people notice absences only relative to what they believe should be present.
The experiment tests two complementary factors that influence this process: the form of reference framing, which determines how expectations are anchored, and attentional guidance, which determines whether participants are prompted to search for absences.

\subsection{Reference Framing}
Building on Gigerenzer et al~\cite{gigerenzer1988presentation}, we propose that the perception of information surplus or absence depends on
how population-level context is represented and experienced. In their experiments, participants correctly incorporated
base-rate information only when they observed the sampling process: they could see an urn containing, for example, 70 engineer cards and 30 lawyer cards and observe one being drawn. This visual experience created
an internal model of the population.  When the same numerical information was only verbally asserted, participants no longer treated it as evidence about the population. Instead, they focused on the descriptive traits of each case (for instance, whether a person sounded like an engineer or a lawyer) and ignored the stated base rates.
We apply this principle to the domain of reference-based data interpretation. The experiment manipulates how reference information, that is, the information used to form expectations about the focal entity, is presented. A partial reference displays several concrete exemplars from a comparison set. In the energy domain, for example, this corresponds to showing the energy production of several other countries. This format parallels Gigerenzer et al.’s ``observable sampling'' condition, where population structure becomes accessible through concrete examples. A global reference, by contrast, summarizes the entire population through an aggregate visualization such as a mean or range. In the energy example, this corresponds to displaying the world’s average energy production profile.

Two types of reference frames are compared:
\begin{description}
    \item[Partial reference ] In this condition, people are presented with several concrete examples from the population as a reference. This framing invites reasoning and active expectation building. This prediction is also consistent with Neisser’s view that partial input initiates an anticipatory schema and recruits exploratory processing to complete what is expected. By requiring participants to infer the broader pattern from exemplars, partial framing should strengthen the internal reference model against which absences can be registered \cite{neisser1976cognition}.

    \item[Global reference] In this condition, people are presented with framing that summarizes the population, supporting abstract, norm-based comparison.
\end{description}
Similar to Gigerenzer et al and Neisser, we expect that constructing expectations from partial samples will heighten sensitivity to what is missing. When viewers must infer the population pattern from limited examples, they engage more deeply in building a mental model of what should be present, making deviations and absences more noticeable.

In contrast, a global statistical reference may promote passive comparison to an abstract average, reducing the salience of specific missing elements.

\textbf{H1 (Reference Framing).} Absence detection will be higher under a \textit{partial reference} than under a \textit{global reference}, because inferring expectations from a few concrete examples makes missing information more salient.

\subsection{Guided Attention}
People naturally attend to what is present. Known as the feature-positive effect~\cite{newman1980feature}, this is a globally acknowledged phenomenon. 
Explicit prompts that direct attention to what might be missing should counter this bias and elicit more systematic search for absence~\cite{tversky1974judgment}.

\textbf{H2 (Guided Attention).} Guided prompts will markedly increase absence detection compared with spontaneous descriptions, regardless of reference framing.

\section{Methodology}
In order to explore absence detection, we conducted a remote study in which 100 participants interpreted bar-chart visualizations comparing a focal entity to either global or exemplar-based references, first spontaneously and then under guided prompting. The study involving human participants was reviewed and approved [Omitted for anonymity] 
Participants were recruited via the Prolific platform, where they reviewed an online consent form and provided informed consent prior to participation. All participants received monetary compensation and could exit at any time without penalty. No personally identifying information was collected.
\subsection{Participants and Design}

\begin{figure}[t]
\centering
\includegraphics[width=.8\linewidth]{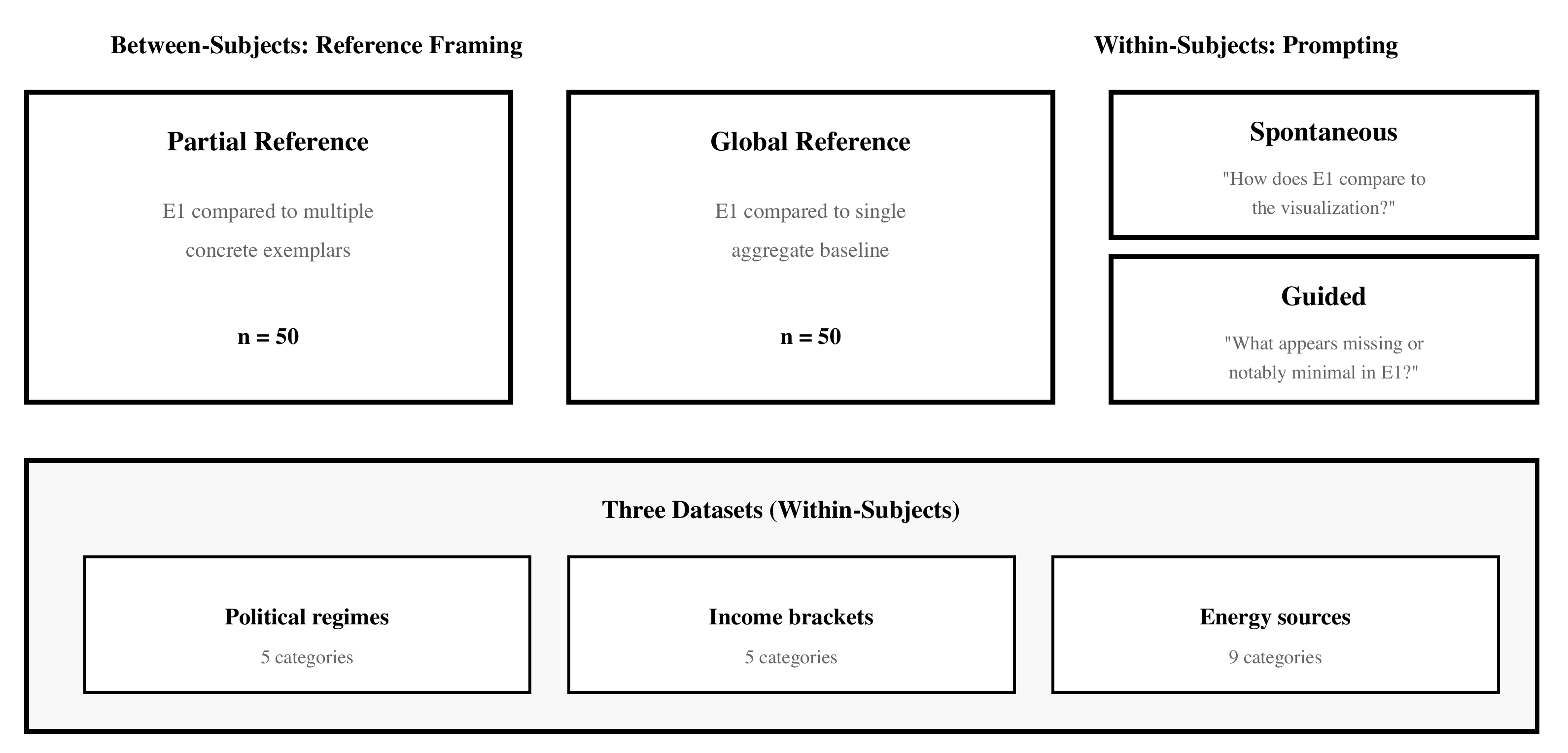}
\caption{Overview of the experimental design. The study employed a 2×2 mixed factorial design with Reference Framing (Partial vs. Global) as between-subjects and Prompting Mode (Spontaneous vs. Guided) as within-subjects. Participants (N = 100) were randomly assigned to one of two reference conditions: Partial Reference, where the focal entity (E1) was compared to multiple concrete exemplars, or Global Reference, where E1 was compared to a single aggregate baseline. All participants responded to both prompting conditions across three datasets (political regimes, income brackets, and energy sources), resulting in six open-ended responses per participant.}
\label{fig:exp_design}
\end{figure}

As illustrated in Figure~\ref{fig:exp_design}, the study followed a mixed factorial design combining between- and within-subject factors. We recruited 100 participants through the Prolific online platform to complete a web-based visualization interpretation task. Participants were randomly assigned to one of two \textit{reference framing} conditions: a \textit{Partial Reference} condition presenting several concrete exemplars and a \textit{Global Reference} condition presenting an aggregated population baseline. Reference framing was manipulated between subjects, while \textit{prompting mode} (Guided vs.\ Spontaneous) and \textit{dataset domain} (Regime, Wealth, Energy) varied within subjects. Each participant viewed three datasets under both prompting conditions and produced a total of six open-ended text responses.

\begin{table}[ht]
\centering
\caption{Participant demographics}
\begin{tabular}{|l|l|}
\hline
\textbf{Demographics} & \textbf{Values} \\ \hline
\textbf{Age} & Range: 22–80; Mean: 42; SD: 13.1 \\ \hline
\textbf{Gender} & Female: 41\%; Male: 59\% \\ \hline
\textbf{Education} & High School: 26\%; Bachelor’s: 41\%; Master’s: 20\%; Other: 13\% \\ \hline
\textbf{Country of Residence} & United States: 62\%; United Kingdom: 34\%; Other: 4\% \\ \hline
\textbf{Ethnicity} & White: 67\%; Asian: 15\%; Black: 4\%; Mixed/Other: 8\%; Unspecified: 6\% \\ \hline
\textbf{Employment Status} & Full-time: 41\%; Part-time: 7\%; Not in paid work: 7\%; Unemployed: 5\%; Other/No answer: 40\% \\ \hline
\end{tabular}
\label{tab:demographics_detailed}
\end{table}

Table~\ref{tab:demographics_detailed} summarizes the demographic composition of the sample. Participants represented a diverse adult population with balanced gender and educational backgrounds. Most held higher education degrees and resided primarily in the United States and the United Kingdom. The two experimental conditions were comparable in demographic makeup.

\subsection{Data and Conditions}

Three datasets captured distinct categorical domains (Table~\ref{tab:datasets}): political regimes, income distribution, and energy production. The visualization presented were as vertical bar charts showing percentage distributions across categories. Each chart included a focal entity labeled \textbf{E1} (highlighted in yellow) compared with one or more reference entities. E1 was constructed as an extreme case concentrated in a single category—representing a strong presence of one feature and notable absence of others—while reference entities displayed more balanced distributions.

\begin{table}[t]
\centering
\caption{Datasets and variables used in the experiment.}
\label{tab:datasets}
\begin{tabular}{p{3.5cm}p{6cm}p{5.5cm}}
\toprule
\textbf{Dataset} & \textbf{Variable Description} & \textbf{Categories} \\
\midrule
\textbf{Regime Type Distribution} & 
Percentage of population living under each political regime type & 
Closed Autocracy, Electoral Autocracy, Democracy, Liberal, No Regime \\
\addlinespace
\textbf{Wealth Distribution} & 
Percentage of population in each daily per-capita income bracket & 
Up to \$3, \$3–\$4.20, \$4.20–\$8.30, \$8.30–\$10, Above \$10 \\
\addlinespace
\textbf{Energy Production} & 
Percentage of annual energy production by source (TWh) & 
Bio, Coal, Gas, Hydro, Nuclear, Oil, Other, Solar, Wind \\
\bottomrule
\end{tabular}
\end{table}

\paragraph{Partial Reference (Concrete Exemplars).}  
E1 appeared alongside two or three specific entities, providing visible samples from the broader population and enabling reasoning by analogy to observed variation. Example comparisons included Asia, North America, and Europe in the \textbf{Regime Type} dataset; China, Indonesia, and Ireland in the \textbf{Wealth} dataset; and Costa Rica, Cuba, and France in the \textbf{Energy} dataset.
\begin{figure}[th]
\centering
\includegraphics[width=0.6\linewidth]{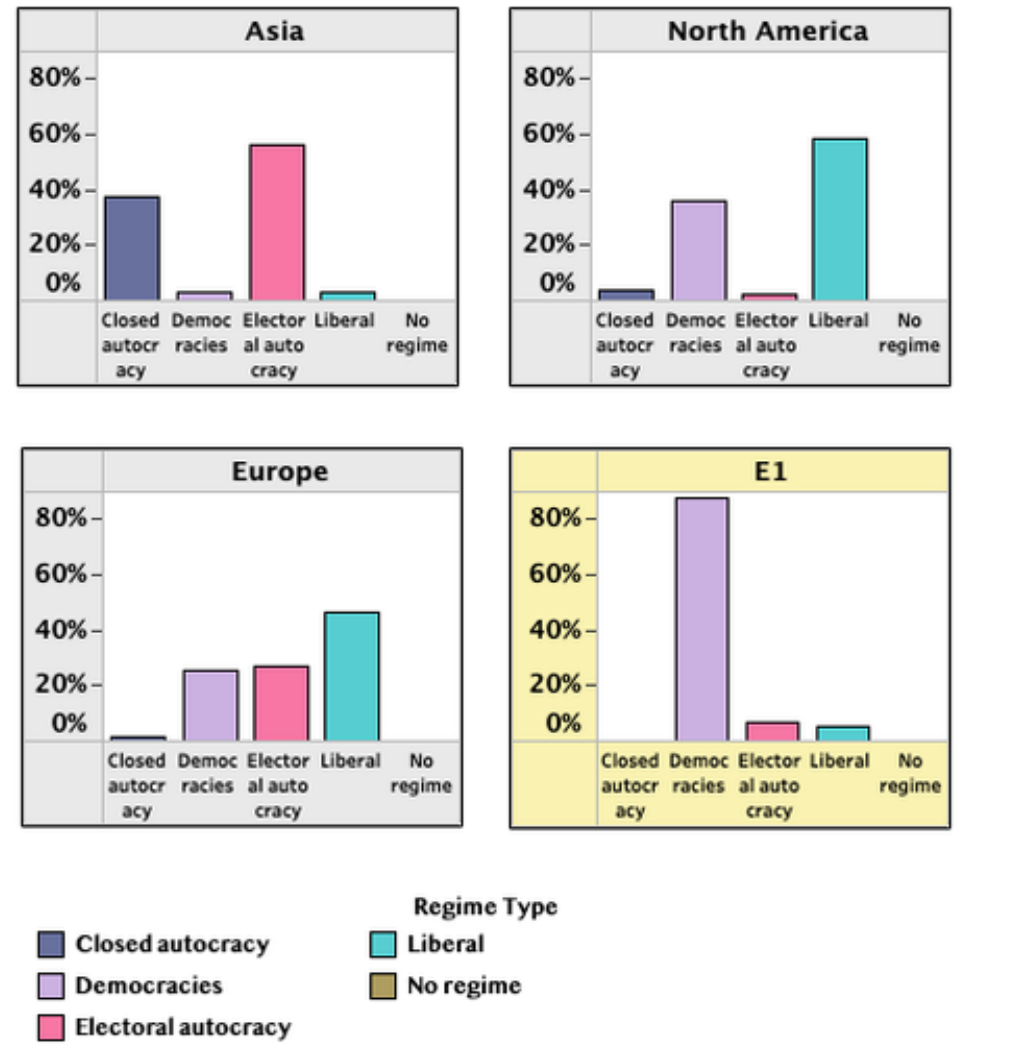}
\caption{Example visualization in the \textit{Partial Reference} condition using the Regime dataset. The focal entity (E1, highlighted in yellow) is compared with three exemplar regions: Asia, North America, and Europe.}
\label{fig:partial_example}
\end{figure}
Figure~\ref{fig:partial_example} illustrates this condition, where the focal entity (E1) is compared with several concrete instances representing distinct distributions.

\paragraph{Global Reference (Summary Aggregate).}  
E1 was displayed alongside a single entity labeled \textbf{World}, representing the aggregated population distribution for the same variable. This reference provided an overall statistical norm rather than individual examples. For instance, in the \textbf{Regime} dataset, the World bar showed the global percentage of populations under each regime type (approximately 45\% electoral autocracy, 35\% democracy, 20\% other). In the \textbf{Wealth} dataset, the global income distribution was centered on the \$4.20–\$8.30 range, and in the \textbf{Energy} dataset, the global energy mix was dominated by coal and gas, with hydro, oil, and renewables.
\begin{figure}[th]
\centering
\includegraphics[width=0.6\linewidth]{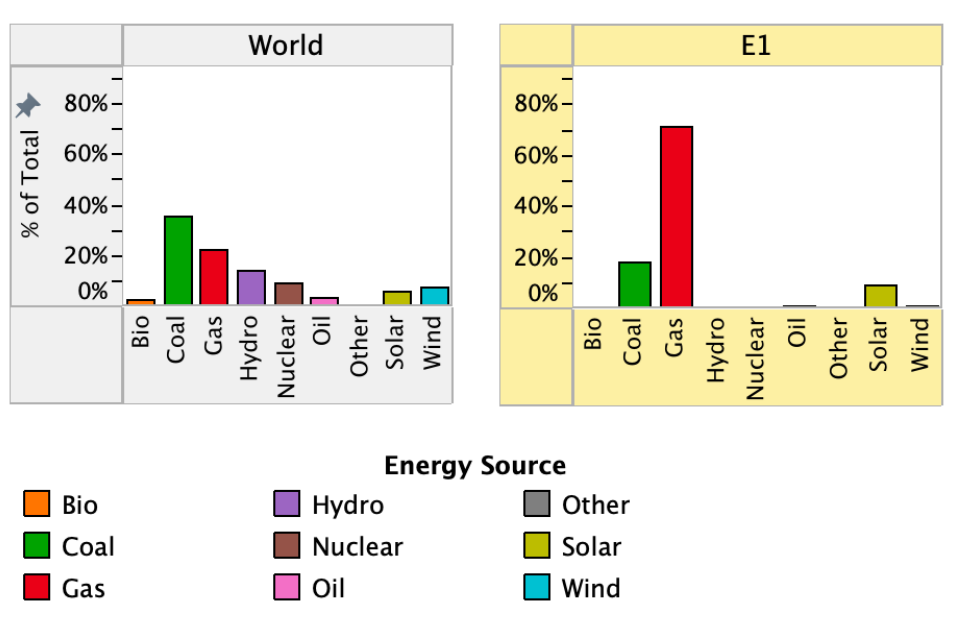}
\caption{Example visualization in the \textit{Global Reference} condition for the same Regime dataset. The focal entity (E1, yellow) is compared with a single World aggregate representing the population distribution across regime types.}
\label{fig:global_example}
\end{figure}
Figure~\ref{fig:global_example} shows this condition, in which E1 is evaluated against a unified global baseline that encapsulates the population-wide expectation.
\subsection{Procedure}

Participants viewed visualizations from the three datasets under their assigned reference condition. In the \textit{spontaneous} phase, they answered the open-ended question:  
\textbf{``How does E1 compare to what you observe in this visualization?''}  
These initial responses captured natural, unprompted comparative reasoning. In the subsequent \textit{guided} phase, the same visualizations were shown again with the targeted prompt:  
\textbf{``What would you expect to see in E1 that appears to be missing or notably minimal?''}  
Separating the two phases ensured that the first responses reflected intuitive impressions unshaped by explicit direction to search for absences.

\subsection{Data Analysis}

To analyze the open-ended responses, we developed two large language models (LLMs) that were integrated into the research workflow: the \textit{Absence Detector} and the \textit{Surplus Detector}. These modules automatically identified linguistic signals of absence or surplus within participants’ textual explanations.
Our module's prompts and code, are available in our 
repository \footnote{\url{ https://anonymous.4open.science/r/IUI_anonymized-DD71/README.md}}
\paragraph{Automated analysis pipeline.}
All responses were exported from Google Sheets, lowercased, and tokenized using the \textsf{tidytext} and \textsf{stringr} packages in~\textsf{R}. 
Stopwords, punctuation, and HTML artefacts were removed prior to model input.  
Each cleaned response was then passed to the \textit{Gemini-2.5-flash} engine using a standardized API call (temperature~=~0.2, top-$p$~=~0.9) to ensure stable, deterministic behavior across runs.  
The model was prompted with the following fixed instruction:

\begin{quote}
``You are an expert in literary and linguistic analysis, specializing in identifying themes of absence, loss, and omission.  
Analyze the given text and determine whether it conveys the idea of `missing' or `absence'.  
Respond only with a JSON object containing a Boolean field (\texttt{"absence": true/false}) and a brief justification string.''
\end{quote}

Each response returned a structured JSON object containing:  
(i)~a binary indicator of whether absence was expressed, and  
(ii)~a short textual explanation of the model’s reasoning.  
For example, the input \textit{``E1 has no closed autocracy''} was classified as \texttt{"absence": true} with the justification \textit{``explicit negation of a category indicates missingness''}.  
All outputs were logged and version-controlled to ensure full reproducibility.


This approach provided a reproducible and linguistically grounded method for classifying nuanced semantic expressions of \textit{absence} in participants’ narratives, aligning closely with human expert judgments~\cite{tang2024harnessing}.  

\paragraph{Validation.}
To verify reliability, a random $20\%$ subset of responses ($n{=}120$) was independently coded by a trained linguist specializing in English semantics.  
Inter-annotator agreement between the human coder and the LLM output reached $\kappa=0.84$ (substantial agreement).  
Based on this validation, the model achieved an overall accuracy of~0.92, precision of~0.94, recall of~0.89, and $F_1$~score of~0.92.  
Discrepancies were reviewed manually to identify systematic errors, which primarily involved ambiguous cases (e.g., implied absence without explicit negation).

All source code, prompts, and validation procedures are archived in the project’s open repository which will be provided upon acceptance.

\paragraph{Statistical analysis.}
Because the dependent measures are binary (detection vs.~no detection), categorical comparisons between conditions were analyzed using Pearson’s $\chi^2$ tests with Yates’ continuity correction.  
Separate tests evaluated (a)~the effect of reference frame within each prompting mode, and (b)~the effect of prompting mode within each reference frame.  
This approach aligns with the factorial design, where both manipulations are nominal variables with two levels each.  
$\chi^2$ tests were chosen because they assess differences in proportions without assuming normality, making them suitable for frequency-based comparisons with discrete outcomes.  

Effect sizes were computed using Cramer’s~$V$~\cite{cramer1999mathematical}, which provides a normalized measure of association strength for contingency tables ($0$–$1$).  
Condition-wise means and standard deviations of detection proportions were calculated using the \textsf{group\_by()} and \textsf{summarise()} functions in~\textsf{dplyr}.  
All quantitative analyses were conducted in~\textsf{R} (version~4.2.0) using the \textsf{dplyr}, \textsf{tidyr}, and \textsf{ggplot2} packages~\cite{wickham2019welcome}.

\section{Results}

\subsection{Absence Detection by Reference Framing and Prompting Mode}
\begin{table}[h]
\centering
\small
\begin{tabular}{@{}l cc ccc cc ccc@{}}
\toprule
 & \multicolumn{5}{c}{\textbf{Spontaneous Condition}} &
   \multicolumn{5}{c}{\textbf{Guided Condition}} \\
\cmidrule(lr){2-6} \cmidrule(lr){7-11}
\textbf{Domain} & \textbf{Global} & \textbf{Partial} &
\textbf{$\chi^2$} & \textbf{$p$} & \textbf{$V$} &
\textbf{Global} & \textbf{Partial} &
\textbf{$\chi^2$} & \textbf{$p$} & \textbf{$V$} \\
 & \textbf{(\%)} & \textbf{(\%)} & \textbf{(1)} & & &
 \textbf{(\%)} & \textbf{(\%)} & \textbf{(1)} & & \\
\midrule
Energy   & 28.0 & 58.0 & 9.18  & $<.01$   & 0.30 &
         88.0 & 94.0 & 1.10 & $\geq.10$ & 0.10 \\
Wealth   & 28.0 & 42.0 & 2.15  & $\geq.10$ & 0.15 &
         84.0 & 90.0 & 0.80 & $\geq.10$ & 0.09 \\
Regime   & 36.0 & 50.0 & 2.00  & $\geq.10$ & 0.14 &
         92.0 & 90.0 & 0.12 & $\geq.10$ & 0.03 \\
Combined & 30.7 & 50.0 & 11.65 & $<.001$  & 0.20 &
         88.0 & 91.3 & 0.90 & $\geq.10$ & 0.05 \\
\bottomrule
\end{tabular}
\caption{Framing Effect Summary: Global vs.\ Partial Reference. Detection rates and statistical comparisons across domains under Spontaneous and Guided conditions.}
\label{tab:framing_effects}
\end{table}

\paragraph{Framing effects.}
Table~\ref{tab:framing_effects}  summarizes how reference framing (Global vs.\ Partial) influenced absence detection under both spontaneous and guided conditions.  
Across domains, reference framing produced a clear pattern in the \textit{spontaneous} condition but not in the \textit{guided} one (Table~\ref{tab:framing_effects}). When participants were not explicitly prompted to look for what was missing, absence detection was substantially higher under the \textit{Partial Reference} framing (mean = 50.0\%) than under the \textit{Global Reference} framing (mean = 30.7\%), $\chi^2(1)=11.65$, $p<.001$, $V=.20$. Results were strongest in the Energy dataset and consistent across the other domains. When attention was guided toward absences, detection rates converged across framings.

\paragraph{Prompting effect. }
\begin{table}[h]
\centering
\small
\begin{tabular}{@{}lcccccc@{}}
\toprule
\multicolumn{7}{c}{\textbf{Prompting Effect Summary: Spontaneous vs.\ Guided}} \\
\midrule
\multicolumn{7}{l}{\textbf{Global Reference}} \\
\textbf{Domain} & \textbf{Spontaneous (\%)} & \textbf{Guided (\%)} & \textbf{Difference (pp)} &
\textbf{$\chi^2(1)$} & \textbf{$p$} & \textbf{$V$} \\
\midrule
Energy   & 28.0 & 88.0 & +60.0 & 36.92 & $<.001$ & 0.61 \\
Wealth   & 28.0 & 84.0 & +56.0 & 32.08 & $<.001$ & 0.57 \\
Regime   & 36.0 & 92.0 & +56.0 & 33.60 & $<.001$ & 0.58 \\
Combined & 30.7 & 88.0 & +57.3 & 102.14 & $<.001$ & 0.58 \\
\midrule
\multicolumn{7}{l}{\textbf{Partial Reference}} \\
\textbf{Domain} & \textbf{Spontaneous (\%)} & \textbf{Guided (\%)} & \textbf{Difference (pp)} &
\textbf{$\chi^2(1)$} & \textbf{$p$} & \textbf{$V$} \\
\midrule
Energy   & 58.0 & 94.0 & +36.0 & 18.69 & $<.001$ & 0.43 \\
Wealth   & 42.0 & 90.0 & +48.0 & 25.53 & $<.001$ & 0.51 \\
Regime   & 50.0 & 90.0 & +40.0 & 19.05 & $<.001$ & 0.44 \\
Combined & 50.0 & 91.3 & +41.3 & 63.11 & $<.001$ & 0.46 \\
\bottomrule
\end{tabular}
\caption{Prompting Effect Summary: Spontaneous vs.\ Guided. Improvements in detection rate and statistical comparisons within each reference frame across domains.}
\label{tab:prompting_effects}
\end{table}
Across all domains and reference conditions, prompting produced a large and highly significant increase in absence detection (Table~\ref{tab:prompting_effects}). When participants were explicitly asked to look for what might be missing, detection rates rose sharply—from roughly one-third of participants in the spontaneous phase to nearly nine in ten under guided prompting. Across both reference types, explicit guidance produced large, highly significant gains ($\approx +40–60$ pp, $V \approx .5–.6$), confirming a robust prompting effect. Large Cramer's $V$ values ($.43$–$.61$) across all tests indicate a strong effect size. 

\subsubsection{Qualitative Example}
\begin{table}[h]
\centering
\caption{Illustrative participant responses across spontaneous and guided conditions.}
\label{tab:absence_detector_results}
\begin{tabular}{p{2.2cm} p{10cm} p{1.6cm}}
\toprule
\textbf{Condition} & \textbf{Participant's Answer} & \textbf{Absence Detected?} \\
\midrule
Spontaneous & E1 is very democratic & False \\[3pt]

Spontaneous & Gas represents 70\% of the production, which is a lot higher than any other country. & False \\[3pt]

Spontaneous & E1 is predominantly democratic (80\%), well above the world average. It has almost no autocracies or authoritarian electoral regimes, suggesting a more stable and participatory political environment. & True \\[3pt]

Spontaneous & E1 produces the most gas, coal and solar, respectively, of all the countries represented. E1 is also the only country that does not produce hydro energy. & True \\[3pt]

Spontaneous & The daily per capita income shown in E1 is mostly over \$10, showing a wealthy country compared to China and Indonesia, but not dissimilar to Ireland, although E1 has a small number of lower earners that Ireland does not & True \\[3pt]

Guided & Data labels, explanation of terms, overall population number & False \\[3pt]

Guided & Nothing appears to be missing. The totals appear to sum to ~100\%. It just contrasts with the other countries given the dependence on gas & False \\ [3pt]

Guided & A lack of eco fuels, namely nuclear, hydro, bio, and wind. & True \\[3pt]

Guided & E1 has no one making \$4.20 or below. & True \\[3pt]

Guided & E1 has far less hydropower but much more gas use than the others & True \\

\bottomrule
\end{tabular}
\end{table}

Table \ref{tab:absence_detector_results} provides illustrative examples of participant responses across spontaneous and guided conditions, with the engine's evaluation. Generally, responses shifted from descriptive commentary in the spontaneous phase to explicit absence language once attention was guided, with markers such as lack, no one, and doesn’t have any. 

\subsection{Patterns Across Datasets}

\begin{figure}[ht]
\centering
\begin{subfigure}[t]{0.9\textwidth}
    \includegraphics[width=\linewidth]{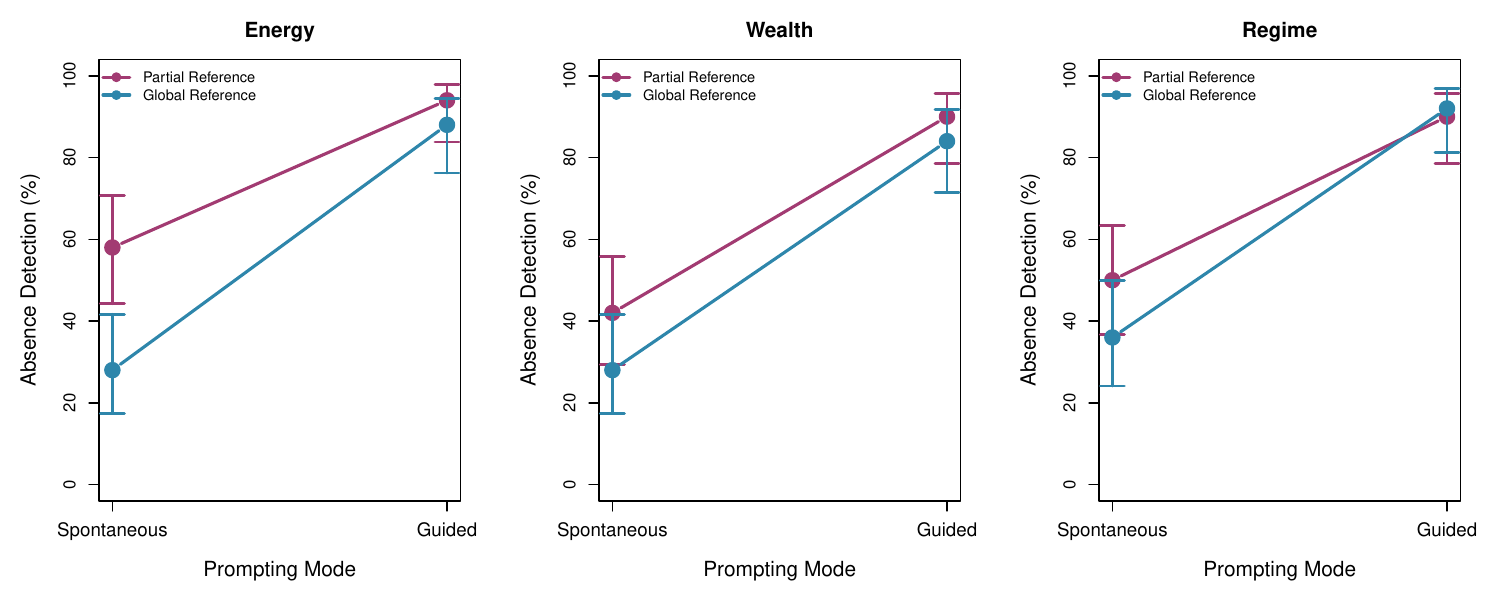}
    \caption{\centering Absence detection rates across the three domains (\textit{Energy}, \textit{Wealth}, and \textit{Regime}), under Spontaneous and Guided prompting and two reference frames (Partial and Global).}
    \label{fig:domain_absence}
\end{subfigure}

\vspace{1em}

\begin{subfigure}[t]{0.5\textwidth}
    \includegraphics[width=\linewidth]{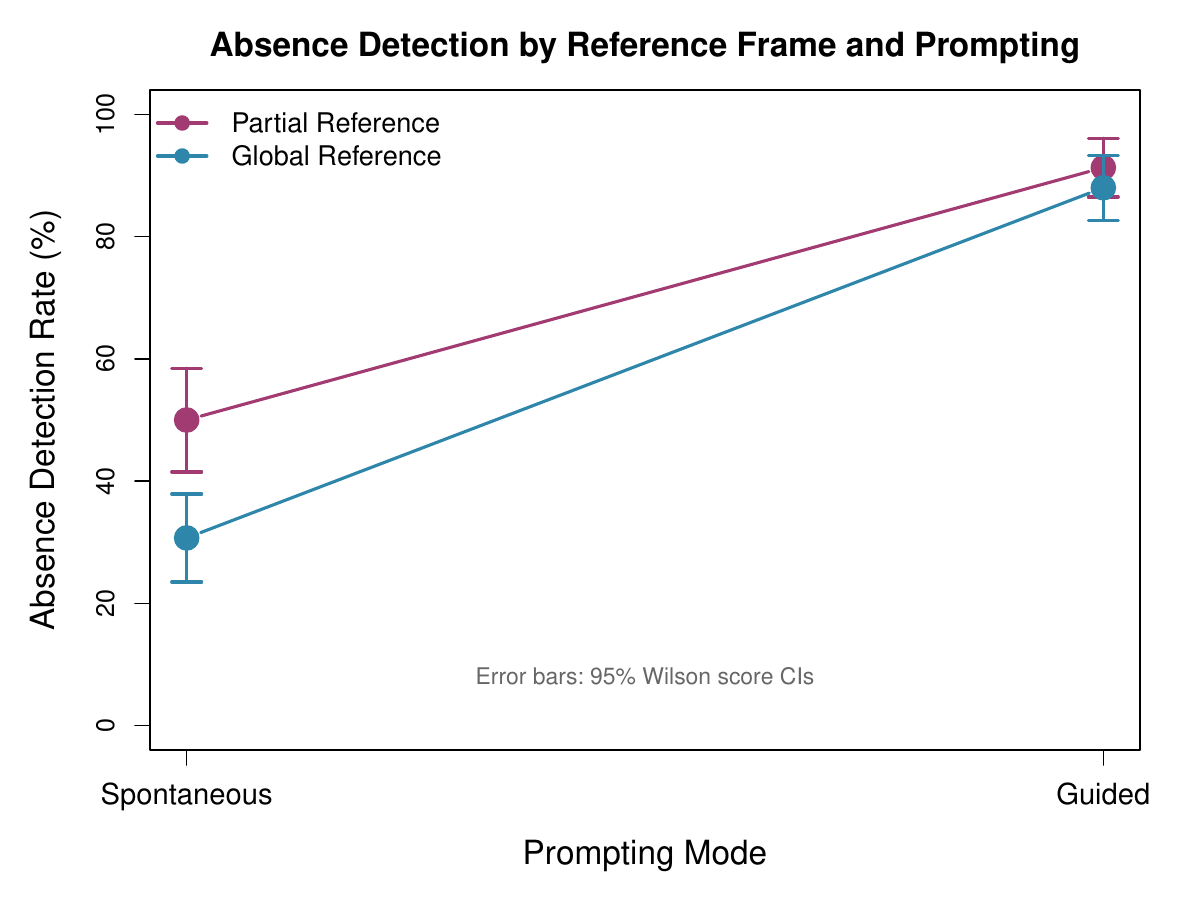}
    \caption{\centering Interaction between prompting mode and reference framing, averaged across domains.}
    \label{fig:overall_absence_interaction}
\end{subfigure}

\caption{\centering Absence detection across reference framing and prompting conditions. 
Panel~(a) shows per-domain detection rates, highlighting sharp increases from spontaneous to guided prompting across all contexts. 
Panel~(b) summarizes the overall interaction, showing that guided prompting eliminates the small spontaneous advantage 
of the Global reference. Error bars denote 95\% Wilson score confidence intervals.}
\label{fig:absence_interactions}
\end{figure}

Figure~\ref{fig:absence_interactions} presents absence detection patterns across experimental conditions. 
Panel~(a) shows detection rates for the three domains (\textit{Energy}, \textit{Wealth}, and \textit{Regime}) under 
Spontaneous and Guided prompting, each compared across Partial and Global reference frames. 
Panel~(b) summarizes the overall interaction between reference framing and prompting mode. 
Error bars in both panels represent 95\% Wilson score confidence intervals.
Across datasets and domains, spontaneous absence detection was low but rose sharply when participants were guided to search for missing information.

These results confirm that absence recognition depends strongly on attentional direction rather than perceptual availability.

\subsection{Summary of Main Findings}

We summarize the results in the following key findings.

\begin{enumerate}

\item \textbf{Moderate spontaneous detection of absence.}  
In the spontaneous phase, when participants freely described what stood out in the visualizations, absence detection was generally low.  
Spontaneous absence detection was generally low ($\approx$ 40\%), showing that most participants focused on visible features even when cues were available.

\item \textbf{Prompting as the dominant driver of absence recognition.}  
When guided, detection nearly doubled across domains ($\approx$ +50 pp), a large, highly significant effect ($V \approx.5$). 
\item \textbf{Reference framing: small but consistent influence.}  
Partial references modestly improved detection compared to Global referencing ($\Delta \approx 10–15$ pp; $V \approx .12$), indicating small but consistent contextual facilitation.
\item \textbf{No strong interaction between framing and prompting.}  

\item \textbf{Cognitive guidance over visual availability.}  
The clear difference between spontaneous and guided phases shows that the difficulty of perceiving absence lies not in visual salience but in cognitive framing.  
Identical visualizations elicited qualitatively different interpretations depending solely on whether participants were prompted to search for what was missing.  
Guidance effectively redirected attention from presence-based description to absence-based reasoning, revealing that absence detection in visualization is governed primarily by how attention is directed, not by what is visibly present.
\end{enumerate}

\paragraph{Statistical Note.}
We conducted four planned comparisons (Partial vs Global, Spontaneous vs Guided), all defined a priori, so no multiple-testing correction was applied~\cite{perneger1998s}. 

Chi-square tests revealed large, highly significant prompting effects 
for both reference types (Global: $\chi^2(1)=102.17$, $p<.001$, $V=0.58$; 
Partial: $\chi^2(1)=61.81$, $p<.001$, $V=0.45$). 
Reference frame effects were smaller: a modest but reliable difference 
in the spontaneous phase ($\chi^2(1)=11.65$, $p=.0006$, $V=0.20$) 
and a non-significant difference in the guided phase 
($\chi^2(1)=0.90$, $p=.343$, $V=0.06$). 
Applying a conservative Bonferroni correction ($\alpha=.0125$) 
does not alter these conclusions: both prompting effects remain highly significant, 
and only the spontaneous framing effect meets the adjusted threshold. 
Fisher’s exact tests yielded equivalent results, confirming robustness to small expected cell sizes.

\section{Discussion}
\subsection{Theoretical Implications}
The present study advances the understanding of how reference framing and cognitive guidance shape human recognition of absence.
Consistent with theories of expectation-based perception~\cite{margoni2024violation}, our findings demonstrate that detecting missing information depends on the availability and explicitness of a reference model.

Across domains, spontaneous absence detection varied substantially by reference condition.
While many participants initially focused on present features, a notable proportion, especially under the \textit{partial reference}, identified missing categories even without explicit prompting.
This shows that the limitation is not an inability to perceive absence but a tendency for attention to default toward visible information unless expectations are actively engaged.

\paragraph{Reference effect.}
The results show that the kind of reference frame provided influences the way people form expectations: a \textit{global frame} supplies an explicit baseline, while a \textit{partial frame} prompts them to infer it from examples.
This distinction seems to differentiate between two cognitive pathways for comparison.
In the global condition, expectation is externally provided. The viewer’s task is to align observations with a ready-made standard.
In the partial condition, expectation must be internally constructed from limited cues, requiring participants to abstract a pattern that generalizes beyond the given examples.
These internally generated expectations are effortful but more cognitively engaging: they invite hypothesis formation and promote active comparison between what is observed and what is missing. Our results indicate that the participants noticed absence more in the partial condition, confirming that encouraging participants to construct their expectations enhances the awareness of absence.

This confirms that reference framing is not a neutral visualization choice but an active determinant of cognitive access to expectation.

Thus, \textbf{visual framing determines whether expectation is supplied by the system or built by the user}.

\paragraph{Guidance effect.}
Guidance amplified absence detection across all conditions: when participants were asked to consider ``what should be there but isn’t'', detection rates rose sharply. This contrast shows that recognition of absence depends less on the visibility of the data and more on whether viewers are encouraged to think about what should be present and notice when it is missing.

This result aligns with the long-documented \textit{feature-positive effect}~\cite{newman1980feature,allan1993human}: organisms and humans detect and learn from presences more readily than from absences.
Neuroscientific evidence further suggests that perceiving ``nothing'' recruits distinct neural systems from those encoding positive features~\cite{nieder2016representing}.
In our task, this asymmetry manifested behaviorally: participants naturally described what they saw rather than what they failed to see.
Yet, the rapid shift following an explicit absence-oriented prompt indicates that absence reasoning is cognitively available but typically dormant, requiring attentional reframing to surface. Notably, even when directly prompted, a few claimed that nothing is missing, or looked for missing details in the graph, e.g., ``E1 has no country name''.

\paragraph{Cognitive load.} 
Absence detection is shaped by heuristics beyond the feature-positive effect and is highly sensitive to cognitive load. Under higher load, viewers are more likely to rely on fast, memory-driven strategies rather than systematically validating which categories should appear. The availability heuristic may bias attention toward categories most salient in memory (or visually prominent), making less familiar or less salient absences harder to notice. Similarly, \emph{representativeness} may lead viewers to treat distributions that resemble familiar prototypes as ``complete,'' prematurely terminating search even when semantically important categories are missing.  In addition, 
partial references require viewers to synthesize expectations from a small set of exemplars. When users possess relevant domain knowledge, such a synthesis enables them to infer what ``should'' be present and to identify when it is not. However, when domain familiarity is low, this inferential process may be more fragile~\cite{gigerenzer1988presentation}. This highlights an important design trade-off: partial framing promotes active expectation construction, yet it may place greater cognitive demands on novice users. These heuristics help explain variability in spontaneous absence detection and illustrate why targeted prompting can be so effective.
Such a synthesis enables them to infer what ``should'' be present and to identify when it is not. However, when domain familiarity is low, this inferential process may be more fragile~\cite{gigerenzer1988presentation}. This highlights an important design trade-off: partial framing promotes active expectation construction, yet it may place greater cognitive demands on novice users. These heuristics help explain variability in spontaneous absence detection and illustrate why targeted prompting can be so effective. Extending Neisser's perceptual-cycle account to the domain of absence detection, we propose that partial input can strengthen schema construction by initiating an anticipatory model while still requiring exploratory sampling to ``fill in'' what is expected. In this view, the additional effort induced by partial references is not merely a cost: it can deepen the internal reference structure that makes absences meaningful and detectable~\cite{neisser1976cognition}

Partial and global frames may impose different cognitive loads. Partial framing increases intrinsic load by requiring users to abstract across exemplars, yet this effort appears to deepen expectation formation and sharpen sensitivity to absences. Global framing reduces cognitive load by providing an explicit baseline, but may encourage more surface-level comparisons. Guided prompting introduces an attentional shift that can reduce the cognitive burden of searching for relevant absences. These results suggest that visualization systems should balance cognitive demands with the benefits of expectation construction, adjusting framing strategies to the user’s expertise and task context.

Taken together, the data support a layered cognitive account.
While spontaneous viewing is largely guided by presence-oriented attention, some absence recognition still emerges when a reference frame provides enough context to evoke implicit expectations.
Guided viewing, in turn, amplifies this latent sensitivity by explicitly directing attention toward what should be present but is not.
Reference framing modulates this interplay: a global reference supplies a clear external model for comparison, whereas a partial reference invites participants to construct one internally—both enabling recognition, though through different routes.
The ability to ``see what is not there'' therefore depends not only on explicit guidance, but also on how strongly the visualization itself supports the activation of expectations that make absence perceptible.

\subsection{Practical and Societal Implications}
The results highlight framing as a supporting factor in how users recognize and engage with what is missing.
This finding has direct implications for the design of visualization systems, analytic dashboards, and intelligent interfaces that aim to support reasoning about incomplete, uncertain, or biased data.
Rather than treating reference information as a fixed background, designers can use it as a cognitive means that determines how expectations are formed and compared.

Our results have direct implications for the design of intelligent user interfaces and AI-supported analytic systems. Many AI tools, such as summarization systems, recommender systems, and LLM-based assistants, provide selective information that can implicitly narrow user expectations. Interfaces that surface what is expected but absent can help mitigate over-trust and reveal blind spots in model outputs. Expectation-oriented framing can be integrated into explanation interfaces through exemplar-based comparisons (partial frames) or aggregate baselines (global frames). Guided prompting, as shown in our study, may further support users in identifying omissions in AI-generated summaries or decision-support dashboards. These mechanisms align with broader goals in explainable AI to help users understand not only what a model shows, but also what it leaves out.

\paragraph{Framing as a design parameter.}
A \textit{global frame}, which makes the reference explicit and comprehensive, provides users with a clear normative context for \textit{rapid evaluation}.
Such framing is effective in domains where consistency and efficiency are needed, such as monitoring dashboards in healthcare or finance, where users must quickly detect deviations from established baselines.
By showing overall distributions, averages, or benchmark indicators alongside individual cases, global framing helps users spot large-scale gaps or anomalies with minimal effort.
However, it may also encourage surface-level verification, e.g., checking whether something conforms, rather than reflective questioning about what might be absent.
In contrast, a \textit{partial frame} encourages the construction of expectations from a limited set of examples.
This design strategy engages users more actively: they must infer ``what should be there'' based on partial cues, leading to deeper \textit{cognitive processing} and increased sensitivity to omissions.
Partial framing may therefore be advantageous in exploratory or educational contexts, such as scientific data analysis, public dashboards explaining social indicators, or learning environments, where engagement and hypothesis formation are more important than speed.
In such cases, withholding some reference information (for instance, showing representative samples instead of complete aggregates) can prompt users to actively reconstruct the missing structure, which our data show enhances absence detection even without explicit guidance.

\paragraph{Adaptive and mixed framing.}
Because global and partial references serve different cognitive purposes, visualization systems can treat framing as an interactive design parameter, rather than a static choice.
Interactive interfaces can allow users to shift dynamically between global and partial references, adjusting the level of explicitness as their understanding evolves.
For example, a climate-change dashboard could begin with region-level global baselines for temperature and emissions, then allow users to explore smaller subsets (partial frames) to infer where data are sparse or missing.
Similarly, a hospital analytics interface might start with global performance benchmarks across departments, then switch to partial contextual views that encourage clinicians to identify overlooked risk factors or unrecorded conditions.

\paragraph{Implications for AI and Intelligent User Interfaces}
Although our empirical stimuli are data visualizations, the core idea we present is detecting \emph{absent expected categories} relative to a reference frame. This feature is generalized to AI-mediated sense-making.  LLM-based systems often produce selective summaries or analyses that omit semantically expected elements (e.g., key topics, entities, or subgroups), and users may fail to notice these omissions without an explicit reference. 
Our results motivate expectation-aware IUI strategies, such as making the underlying reference frame explicit (global or partial) and integrating guided prompting as an explanation mechanism to steer attention toward what should be checked. 

\paragraph{Societal Implications}
Through the lenses of framing and expectations, absence detection can inform the design of intelligent interfaces that promote awareness, accountability, and critical engagement with data. By embedding mechanisms that surface missing categories, silent model features, or unreported metrics, designers can turn absences into signals for reflection rather than blind spots. Partial references, as shown in our study, may serve as a useful design strategy to stimulate cognitive attention, inviting users to actively reconstruct what might be missing instead of passively accepting what is shown. Such expectation-oriented framing can help users recognize incompleteness in algorithmic explanations, dashboards, or public data displays, supporting more cautious and equitable interpretations. In this way, interfaces that make absences perceptible do not merely fill information gaps. They help cultivate a mindset of mindful inquiry, strengthening trust and accountability in human–AI collaboration.  This effect is evident in applied domains where absence has been made explicit. Recent work on absence-aware summarization shows that surfacing expected but missing aspects, rather than focusing solely on mentioned content, leads to more informative summaries and supports improved decision making~\cite{fainman2026discomakingabsencevisible}.

\subsection{Limitations and Future Work}
This study has several limitations.
The guided prompt focused only on absences at the second phase. Future work should test symmetric prompts for surpluses.
The task used static bar charts and written responses, unlike real interactive settings.
Automated text analysis may have missed subtle expressions of omission, and the sample of English-speaking adults viewing categorical data limits generalization across cultures and data types.

Future research should further test how expectation framing interacts with domain familiarity, data sparsity, and cognitive load measurement.
A key question emerging from both human and algorithmic evidence is whether systems should aim to help users \textit{construct} expectations or \textit{provide} them explicitly.
Bridging these two modes of internal inference and external reference might advance the design of visualization systems and intelligent interfaces that not only display data, but also help people perceive and reason about what is missing.

\section{Conclusion}
This study shows that people’s ability to recognize what is missing in data depends on how reference information is framed.
When expectations had to be inferred from a few examples (\textit{partial reference}), participants often detected absences as well as—or even better than—those viewing an explicit global baseline.
This suggests that constructing expectations internally can enhance engagement and sharpen sensitivity to missing information.
Under guided prompting, absence detection rose sharply across all conditions, eliminating differences between reference types and confirming that explicit attentional direction can override structural framing effects.
Together, these results indicate that recognizing absence is not a perceptual limitation but an attentional and contextual process shaped by how expectations are built and applied.
Designing visualizations that help users ``see what is not there'' therefore requires cognitive designs and interaction designs that balance between explicit guidance and opportunities for active expectation construction that require deeper attention.

\section*{Generative AI Usage Disclosure}
Generative AI tools (ChatGPT, GPT-5) were used exclusively for language refinement of this manuscript, including minor edits to improve grammar, clarity, and readability. All conceptual development, data analysis, and methodological decisions were made by the authors.
\bibliographystyle{ACM-Reference-Format}
\bibliography{cites}

\appendix

\end{document}